\documentclass[article,preprint]{revtex4}

\usepackage{graphicx}
\usepackage{color}
\usepackage{soul}

\begin{document}

\title{
  Dynamic Control of Topological Defects in Artificial Colloidal Ice
} 
\author{
A. Lib\'{a}l$^{1,2}$, C. Nisoli$^{2}$ C. Reichhardt$^{2}$ and C. J. Olson Reichhardt$^{2,*}$ 
} 
\affiliation{$^1$Faculty of Mathematics and Computer Science, Babe\c{s}-Bolyai University,
  Cluj, 400084, Romania\\
  $^2$Theoretical Division and Center for Nonlinear Studies,
Los Alamos National Laboratory, Los Alamos, New Mexico 87545, USA
} 

\date{\today}
\begin{abstract}

We demonstrate the use of an external field to stabilize and control defect lines connecting topological monopoles in spin ice. For definiteness we perform Brownian dynamics simulations with realistic units mimicking  experimentally realized artificial colloidal spin ice systems, and show how defect lines can grow, shrink or move under the action of direct and alternating fields.  Asymmetric alternating biasing fields can cause the defect line to ratchet in either direction, making it possible to precisely position the line at a desired location.  Such manipulation could be employed to achieve fast, dense, and mobile information storage in these metamaterials.
       
\end{abstract}
\pacs{82.70.Dd,75.10.Hk,75.10.Nr}
\maketitle

Systems mimicking the behavior of spin ice have been studied experimentally
and theoretically for nanomagnetic islands 
\cite{Wang2006,N1,N2,N3,N4,N5,N6,Nisoli2013colloquium,N7,N8,Wang962}, superconducting vortices \cite{Latimer2013,Libal2006,N10}, and 
superparamagnetic colloidal particles on photolitographically etched surfaces \cite{ortiz2016engineering,Libal2006,N11,N12}. In each of these particle-based artificial ice systems, the
collective
lowest energy state is embedded into an ice-manifold where all vertices obey the ``2-in/2-out'' ice rule: two particles are close to each vertex
and two are far from it.
It is possible to write information into such a manifold by using an MFM tip~\cite{Wang962} or an optical tweezer~\cite{ortiz2016engineering} to generate topological defects in the ground state arrangement of the spins.  These defects consist of vertices that violate the ice rule and correspond  to 3-in/1-out or 3-out/1-in configurations.  In magnetic spin ices, such defects are called magnetic monopoles  \cite{Castelnovo2008}. 
In colloidal artificial ice, the defects are not magnetically charged but they still carry a topological charge~\cite{Nisoli2014NJoP}. This implies that they can only appear in pairs separated by a line
of polarized ice-rule vertices,
and disappear by mutual annihilation. In a square ice geometry, such defect lines
are themselves excitations and thus possess
a tensile strength~\cite{Nascimento2012,PhysRevLett.116.077202} that linearly confines the topological charges and
can drive
them to mutual annihilation, restoring the ground state configuration. 

In this paper we show how an additional biasing field can be used to stabilize, control, and move defect
lines
written on the ordered ground state of a square
colloidal artificial spin ice system. To make contact with
recent
experimental realizations of this system
~\cite{ortiz2016engineering,tierno2016geometric},
we employ a gravitational bias
that can be implemented experimentally by tilting the
effectively two-dimensional (2D)
sample.
We consider the interplay of
two completely separate control parameters: the tilt that controls the biasing and
the
perpendicular magnetic field that controls the inter-particle repulsive magnetic forces, as in Ref.~\cite{ortiz2016engineering}.
Adjusting these parameters
gives us precise control over the energetics
of the system and makes it possible to control
the speed of the shrinking or expansion of a defect line.
Then,
using asymmetrical ac biasing fields and taking advantage of the different mobility of the 1-in and 3-in defects in colloidal ice, we show that the defect
line can be made to ratchet, or undergo a net dc motion, in the direction of either of its ends.

The control introduced by the biasing field permits locally stored, compact information to be written into the artificial ice metamaterial by a globally applied field, making it possible to create very dense information storage since the write/read heads need to be situated only at the edge of the memory block. Also, by moving localized packets of information with a global field, it is possible to parallelize the information storage and retrieval procedures, increasing the speed in both cases. 

\subsection*{Results}

\begin{figure}
\includegraphics[width=\columnwidth]{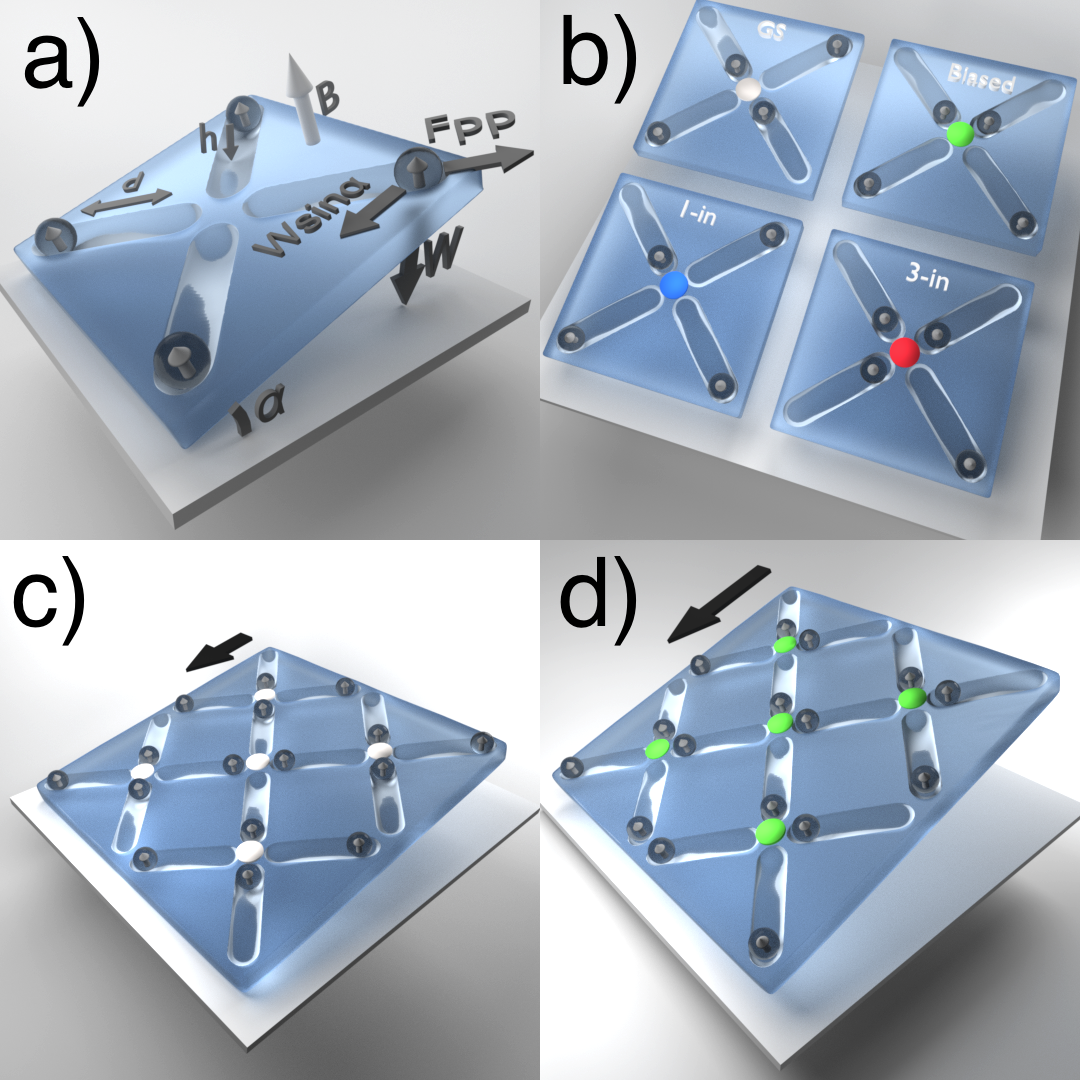}
\caption{
  Schematics of the system. (a) A single vertex
  is surrounded by four double-well pinning sites.
  Labels indicate the distance $d$ between the minima, the barrier height $h$,
  the biasing tilt angle $\alpha$,
  the magnetic field ${\bf B}$, the magnetization ${\bf m}$ it induces in the superparamagnetic particles, and the pairwise magnetic repulsive forces ${\bf F}_{pp}$ acting in
  the sample plane.
  $W$ is the weight of the particle, and the tangential
  component $W\sin (\alpha )$ serves as a biasing force.
  (b) Illustration of four possible vertex arrangements with a
  nonphysical color placed at the vertex
  center to indicate the vertex type.  Ground state (GS, gray), biased
  state (green), 1-in state (blue), and 3-in state (red).  
  (c) The unbiased ground state (gray)
  in a small segment of the sample for a small bias $\alpha$.  (d) The biased
  ground state (green) in a small segment of the sample for a large bias $\alpha$.
}
\label{fig:1}
\end{figure}

{\bf Model and its simulation}
In Figure~\ref{fig:1}, we show schematics of our system illustrating
the interplay between the interparticle and biasing forces.
The four pinning sites
in Figure 1(a)
represent
photolitographically etched grooves in the surface,
each of which acts as a gravitational double well
with a distance of $d=10\mu$m between the two minima.
At the center of the pinning site is
a barrier of height $h=0.87 \mu$m.
Four superparamagnetic colloidal particles are each trapped
in the gravitational wells by the combination of their
own apparent weight ($W =(\rho - \rho_{\rm liquid}) g V $) and the normal force from the wall,
where $\rho$ is the density and $V$ is the volume of an individual particle,
$\rho_{\rm liquid}$ is the density
of the surrounding liquid, and $g$ is the gravitational constant.
A biasing field is introduced
by tilting the whole ensemble by $\alpha$ degrees with respect to the horizontal.
This creates a biasing force
$W\sin(\alpha)$ equal to the tangential projection of the apparent weight of the particles,
providing us with two independent external tuning parameters: the tilt of the surface and the external magnetic field.

The direction of the external magnetic field $\vec B$ is indicated by
a light arrow in Figure 1(a). This field is always perpendicular to the
sample plane, and it induces magnetization vectors $ \vec m\propto \vec B$ parallel to itself in each of the superparamagnetic particles.
As a result, the particles repel each other with an isotropic force $F_{pp} \propto B^2/r^4$ that acts in the plane. This favors arrangements in which the particles maximize their distance from each other. For an isolated vertex the lowest energy configuration is
the 4-out arrangement shown in Figure 1(a); however,
in a system of many coupled vertices,
such an arrangement places an occupancy burden on the neighboring vertices.
As a result, a multiple-vertex arrangement stabilizes in the low energy
ice-rule obeying state
illustrated in Figure 1(c) that is composed of
2-in and 2-out ground state vertices.
The four vertex types we observe are shown in Figure 1(b), where the ground
state vertex is colored gray, the biased
ice-rule obeying vertex is green, the 1-in vertex is blue,
and the 3-in vertex is red.
The 1-in and 3-in monopole
states carry an extra magnetic charge and serve as
the starting and termination vertices for defect lines.
It is also possible for
0-in [Figure 1(a)] and 4-in (not shown) vertices to form,
but they are highly energetically unfavorable and do not play a role in our
defect line study.
For small bias (small $\alpha$),
the ground state
vertex arrangement of Figure 1(c) is favored,
while for large enough $\alpha$, the system switches to the
biased 2-in/2-out arrangement shown in
Figure 1(d).

{\bf Defect line motion.}
Using a $50 \times 50$ vertex
square spin ice sample containing 5000 pinning sites and particles,
we initialize the system
in the ground state by placing the particles inside the appropriate substrate minima.
We then perturb this ground state by introducing 
a defect line to it, achieved by
flipping the effective spins along a diagonal line connecting neighboring vertices.
The defect line is composed of a pair of 1-in and 3-in vertices
connected by a series of biased ground state vertices.
All four possible orientations of the defect line are illustrated in Figure 2(a).
We focus on the dynamics of the defect line in the center of the panel;
all other lines show the same behavior when the biasing field is rotated appropriately. 

In Figure 2(b), to illustrate the contraction of the defect line in the absence of a biasing field when magnetic fields $B$ of different strengths are applied to the system, we plot
the positions $R_1=\sqrt{x_1^2+y_1^2}$ and $R_3=\sqrt{x_3^2+y_3^2}$
of the 1-in and 3-in vertices, respectively,
as a function of time.
We can distinguish several stages of the contraction process.
For very low $B$, particle-particle interactions are very weak
and the defect line remains static, as shown by the constant values of $R_1$ and $R_3$
for
$B=10$ mT.
For small fields in the range of 12 mT $< B <$ 16 mT, the 3-in end of the defect contracts
while the 1-in end of the defect remains static, as shown for $B=12$ and 14 mT.
For $16$ mT $\leq B \leq$ 18.5 mT, both ends of the defect contract, as illustrated
for $B=$ 16, 17, 18, and 18.5 mT.
For $B>18.5$ mT, 
the defect line cannot contract as fast as the rate dictated by the field,
and as a result the line breaks up into 
1-in/3-in vertex pairs along its length.
In a narrow range of fields just above 18.5 mT,
pair formation occurs only near the lower mobility 1-in end of the defect line, since
only this end of the line cannot keep up with the contraction speed.
At slightly higher fields, the 3-in end of the defect line also lags behind the contraction
speed and
nucleation of 1-in/3-in pairs occurs along the whole length of the line.
The nucleation events appear as sudden large jumps in $R_1$ and $R_3$
in Figure 2(b), which arise
when the defect line shrinks by eliminating one or more of the small lines into which 
it has broken instead of by a step-by-step contraction along its length.
We determine the velocity $v_{1(3)}$ of the two defect ends from a linear fit of the
$R_{1(3)}(t)$ curves,
and plot $v_1$ and $v_3$ versus $B$ in Figure 2(c).
For $B<15$ mT only the 3-in end moves, as illustrated in Figure 2(d).
Both ends are mobile
for 15 mT $\leq B \leq 18.5$ mT, but
$v_3>v_1$,
as shown in Figure 2(e).
For $B>18.5$ mT, defect line fracturing and spontaneous 1-in/3-in pair creation along the
defect line occur, as illustrated in Figure 2(f). 

\begin{table}[b!]
\begin{tabular}{|c|c|c|}
\hline
Vertex Type & Particle Configuration & Energy $[10^{-18} J]$ \\
\hline
$0$-in & 0 0 0 0 ($\times$1) & 10.007 \\
$1$-in & 0 0 0 1  ($\times$4) & 15.568 \\
ground state  & 0 1 0 1 ($\times$2) & 24.727 \\
biased $2$-in & 0 0 1 1  ($\times$4) & 32.905 \\
$3$-in & 0 1 1 1  ($\times$4) & 53.837 \\
$4$-in & 1 1 1 1  ($\times$1) & 86.542 \\
\hline
\end{tabular}
\caption{Magnetostatic energy for each vertex type at $B=16$ mT. 
An example configuration for each vertex is
listed. ``1'' (``0'') indicates a colloid close to (far from) the vertex and ($\times n$) indicates
that $n$ different equivalent configurations can be obtained by rotation.
}
\end{table}

Naively one would expect both ends of the defect line to have the same mobility,
$v_1=v_3$, as occurs in
magnetic spin ices. To understand the difference between
$v_1$ and $v_3$, note that although in magnetic spin ice the 1-in and 3-in vertices
have the same energy, in colloidal spin ice they do not.
In dipolar magnetic artificial spin ice~\cite{Nisoli2013colloquium},
frustration occurs at the vertex level and
consists of a frustration of the pairwise interaction.
In contrast, in colloidal spin ice the frustration is a collective effect
arising from the fact that topological charge conservation prevents
vertices from adopting the lowest single-vertex energy
configurations, the 0-in or 1-in states
~\cite{Nisoli2014NJoP}. Thus {\it the colloidal ice-manifold is composed
  of vertices  that are not, by themselves, the lowest energy vertices,
  yet that produce the lowest energy manifold}~\cite{Nisoli2014NJoP}.

To illustrate this point, in Table 1 we list the energy
of each possible vertex configuration in our colloidal spin ice at an external field of $B = 16$mT. 
The table shows that for
the defect line to shrink by moving its 1-in end,
the 1-in vertex must undergo an energetically
unfavorable transformation into a ground state vertex while a biased
vertex makes an energetically favorable transformation into a 1-in vertex.
In contrast, when the 3-in end moves,
a 3-in vertex undergoes an energetically
favorable transformation into a ground state vertex while 
a biased vertex makes an energetically unfavorable transition to a 3-in vertex.
The total energy gain is equal to the transformation
energy of changing a biased vertex into a ground state vertex in each case,
but the initiating transition is energetically favorable for the 3-in end and
unfavorable for the 1-in end,
so that $v_3>v_1$.

Taking into account the overdamped dynamics of the system,
the mechanism for the asymmetry in $v_1$ and $v_3$ 
can be understood more clearly by considering the forces acting on an
individual particle.
During the transition of the particle
from one trap minimum to the other,
both the local force, given by $F_{pp}$ in Eq.~(1), and the substrate force,
given by $F_s$ in Eq.~(1), depend on the position $r_{||}$ of the particle in the trap.
A contraction of the defect line can occur when the local force
is large enough to overcome the substrate force,
$F_{pp}>F_s$.
For simplicity, consider $F^c_{pp,1}$ and $F^c_{pp,3}$, which are the projections of
the local forces acting on the particle parallel to the trap axis
for contraction of the defect line at the 1-in
or 3-in end, respectively.
As shown schematically in Fig.~\ref{fig:3extra},
it is clear that
at the beginning of the switching transition,
$F^c_{pp,3}\approx 2F^c_{pp,1}$ since
the repulsive force acting on the switching particle
is produced by two particles at the 3-in
end but by only one particle at the 1-in end.
As a result,
$v_3>v_1$.

The local forces $F^c_{pp,\alpha}$,
where $\alpha=1,3$, depend quadratically on the applied magnetic field,
allowing us to write $F_{pp,\alpha}=k^c_{\alpha} B^2$
with $k^c_1<k^c_3$.
For small enough $F^c_{pp,\alpha}$,
there is a position
$\bar{r}_{||}$
at which $F^c_{pp,\alpha}(\bar{r}_{||})<F_s(\bar{r}_{||})$.
Writing $F_{\rm trap}=F_s(\bar{r}_{||})$, we see that
when $B$ is small enough, both $F_{pp,1}$ and $F_{pp,3}$
are smaller than $F_{\mathrm{trap}}$ and
$v_1=v_3=0$,
giving a stable (S)
defect line.
When $F_{\mathrm{trap}}/k^c_3<B^2<F_{\mathrm{trap}}/k^c_1$,
$v_3>0$ but $v_1=0$
as in Fig.~2(d), producing a one-sided slow contraction (SC3) state.
For $B^2>F_{\mathrm{trap}}/k^c_1$,
$v_1>0$ and $v_3>0$
as in Fig. 2(e), giving two-sided slow contraction (SC).
There is an even higher critical value for $B$ above which the
local forces acting on the particles {\it within the defect line}
exceed
$F_{\mathrm{trap}}$, permitting the line to disintegrate
via the nucleation of monopole-antimonopole couples.

{\bf Effect of biasing force.}
If we apply a biasing force $F_b$ along a diagonal direction,
as shown in Fig.~1(a), we can change the energy balance between
the ground state and biased ground state vertices.
At sufficiently large $F_b=F_b^0$, 
the biased and ground state vertices have the same energy
so the defect line is stable and does not contract.
For $F_b>F_b^0$,
the biased state becomes energetically more favorable
than the ground state and the defect line begins to grow.
A very high biasing force causes defect lines to nucleate spontaneously
and spread throughout the system until every vertex has switched to the biased state.
In Figure 4(a) we plot the time-dependent position of
the 1-in and 3-in ends of a defect line at different biasing fields.
For high $F_b$, we find a fast contraction (FC) in which,
in addition to the contraction of the line at each end,
we observe spontaneous nucleation of 1-in/3-in vertex pairs along the
line that speed up the contraction.
At very large $|F_b|$,
we observe a global nucleation (GN) of 1-in/3-in pairs that spontaneously produce
defect lines in the bulk
which
propagate through the system until the entire sample reaches
a biased ground state.
In Figure 4(b) we quantify the line contraction by
plotting the total number $N_{\rm biased}$ of biased
ground state vertices in the system.
This measure shows the shrinking, stabilization, and growth of defect
lines for different biasing fields, and can also capture the behavior of the system
when spontaneous nucleation comes into play,
either along the defect line in the case of fast contraction,
or everywhere in the sample in the GN regime.

The interplay between the particle-particle interactions and the biasing field
produces a rich phase diagram, shown in Fig.~4(c) as a function
of $F_b$ versus $B$.
Consider the effects of $F^c_{pp,1}$ and $F^c_{pp,3}$
in the presence of a stabilizing biasing field $F_b$.
The 1-in end is stabilized when $F_b>F^c_{pp,1}-F_{\mathrm{trap}}$.
Thus, the SC3-SC transition follows the line
$F_b=k^c_1B^2-F_{\mathrm{trap}}$.
Similarly, the 3-in end
is stabilized when  $F_b>F^c_{pp,3}-F_{\mathrm{trap}}$, so
the SC3-S transition can be described by
$F_b=k^c_3B^2-F_{\mathrm{trap}}$,
keeping in mind that $k^c_1<k^c_3$.  

If $F_b$ is large enough, rather than merely stabilizing the defect line it can cause the
line to grow.
Figure 4(c) shows
regimes of one-sided slow expansion
(SE3)
on only the 3-in end, as well as slow expansion (SE) on both ends of the string.
We introduce $F^e_{pp,3}=k^e_3 B^2$ and $F^e_{pp,1}=k^e_1 B^2$, which are
the  forces acting on the particles that drive the {\it extension}
rather than the contraction of the 3-in and 1-in ends, respectively.
An elongation of the defect line on the 3-in side
occurs when $F_b>F_{\mathrm{trap}}-F^e_{pp,3}$, so that
$F_b=F_{\mathrm{trap}}-k^e_3 B^2$ describes the S-SE3 transition.
Similarly,  $F_b=F_{\mathrm{trap}}-k^e_3 B^2$ describes the SE3-SE transition line.
For extreme values of $F_b$ in Figure 4(c), 
the biasing field is so strong that the behavior cannot be described
in terms of one-body motion.
Instead, 
the whole sample switches to the biased state by
global nucleation of 1-in/3-in vertex pairs and the spreading of defect lines (GN).

{\bf Ratchet motion under an ac bias.}
By tilting the sample back and forth
over an appropriate range of angles, we can generate an ac external biasing field that
causes the defect lines to oscillate by repeatedly growing and shrinking.
If we allow the biasing field to switch instantaneously,
or at least faster than typical defect speeds, between values $B_a$ and $B_b$,
we can select pairs of biasing fields ($B_a$, $B_b$) for which $v_1 \neq v_3$,
permitting the creation of
a ratchet effect.
In Figure 5 we show
$R_1$ and $R_3$ versus time
under an alternating field where $B_a$ is applied for
$\tau_a=50$ s and $B_b$ is applied for $\tau_b=250$ s per cycle.
Here the defect line
ratchets in the direction of the 3-in end through
a wriggling motion that is composed of
two simple phases.
The field $B_a$ places the sample in the SC regime
where both ends of the line contract with $v_3>v_1$.
Then, under the field $B_b$, the sample enters the SE3 regime
where the line expands only on the 3-in end.
As a result, over successive field cycles
the entire defect line translates in the direction of its 3-in end.
It is also possible to choose the biasing fields in such a way that
under $B_a$ the sample is in the SE regime, where both ends expand with $v_3>v_1$,
while under $B_b$ contraction occurs at only the
3-in end in the SC3 regime.  Under these conditions, the defect line translates
in the direction of its 1-in end, as shown in Figure 5(b).
By adjusting the timing of the expansion and shrinking drives
($\tau_a$ and $\tau_b$), we can slowly shrink, grow or maintain a constant defect line length
as the line ratchets. This makes it possible to re-position
defect segments inside the sample by varying an applied uniform external field.

\subsection*{Discussion}
We have shown that a defect line in a colloidal spin ice system contracts spontaneously
at a rate which increases as the colloid-colloid interaction strength is increased.
The line can be stabilized
by the addition of a uniform global biasing field.
It is possible to control the length and the position of the defect line
by cycling this field to
create oscillations and defect movement through a ratchet effect.
The ratcheting allows us to reposition defect line segments inside the sample to
desired locations after nucleating them at the sample edge,
making it possible to write information into the spin ice
and possibly create a very dense information storage unit.
If the uniform spin ice lattice were replaced by a specifically tailored landscape,
it
is possible to imagine the creation of
logic gates and fan-out positions
where defect lines can merge or split.
Thus it could be possible to construct a device capable of storing and
manipulating the information described by these defect lines
through the creation of ``defectronics'' in spin ice that
could be
the focus of a future study building on defect line mobility and control in spin ices.
Although we concentrate on magnetic colloidal particles, our results could also
be applied to charge-stabilized colloidal systems with Yukawa interactions, for
which it is possible to create large scale optical trapping arrays \cite{R1,R2} and
double-well traps \cite{R3}, and where biasing could be introduced by means of
an applied electric field \cite{R4}.

\subsection*{Methods} 

{\bf Numerical simulation details}
Using Brownian dynamics,
we simulate an experimentally feasible system \cite{ortiz2016engineering} of superparamagnetic colloids placed on an etched substrate of pinning sites.
The spherical, monodisperse particles have a radius of $R = 5.15\mu$m,
a volume of $V = 572.15 \mu$m$^3$ and a density of
$\rho = 1.9 \times 10^3$ kg/m$^3$. They are suspended in water, giving them a relative weight of $W = 5.0515$ pN.
Gravity serves as a pinning force
for the particles placed in the etched double-well pinning sites and also
generates a uniform biasing force $F_b = W \sin(\alpha)$ on all particles
when the
entire sample is tilted by $\alpha$ degrees.  Typically, $\alpha \sim 10^{\circ}$.
The double well pinning sites [Figure 1(a)]
representing the spins in the spin ice are etched into the substrate in the
2D square spin ice configuration [Figure 1(c,d)] with an interwell spacing of $a=29\mu$m.
Each pinning site contains two minima that are $d=10 \mu$m apart.
We place one particle in each pinning site,  which can be achieved experimentally
by using an optical tweezer to position individual particles.
The pinning force $F_s$ acting on the particle is represented by a spring
that is linearly dependent on the distance from the minimum,
so that $F_{s\perp} = 2kW \Delta r_{\perp}$, where $k=1.2 \times 10^{-4}$ nm$^{-1}$ is the spring constant, and $\Delta r_{\perp}$ is the perpendicular distance from the particle to the line connecting the two minima.
When the particle is inside one of the minima,
$F_{s||} = 2kW \Delta r_{||}$, where $\Delta r_{||}$ is the distance from the particle
to the closest minimum along the line connecting them,
while when the particle is between the minima,
$F_{s||} = 8h/d^2 W \Delta r_{||}$, where $h=0.87 \mu$m is the magnitude of the
barrier separating the minima and $\Delta r_{||}$ is the distance between the
particle and the barrier maximum parallel to the line connecting the two minima.
During the simulation, the particles are always attached with these spring
forces to their original pinning sites.

The inter-particle repulsive interaction arises from the magnetization induced by the external magnetic field that is applied perpendicular to the pinning site plane. Each particle acquires a magnetization of $m = B \chi V / \mu_0$, where $B$ is the magnetic field in the range of 0 to 30 mT, $\chi = 0.061$ is the magnetic susceptibility of the particles, and $\mu_0 = 4\pi \times 10^5$pN/A$^2$ is the magnetic permeability of vacuum. The repulsive force between particles is given by $F_{pp} = 3 \mu_0 m^2 / ( 2\pi r^4 ) $, and since it has a $1/r^4$ dependence in a 2D system we can safely cut it off at finite range. We choose a very conservative  cutoff distance of $r_c = 60 \mu$m to include next-nearest neighbor interactions (even though they are negligibly small). 

During the simulation we solve the discretized Brownian dynamics equation:
\begin{equation}
\frac{1}{\mu} \frac{ \Delta x_i} {\Delta t} = 
\sqrt{\frac{2}{D\Delta t}} k_BT N[0,1] + F_{pp}^i + F_s^i + F_b^i 
\end{equation}
where $F_{pp}$, $F_s$ and $F_b$ are the previously described particle-particle,
particle-substrate, and biasing forces, $k_BT = 4.047371$ pN $\cdot$ nm is the thermal energy, $D = 7000$ nm$^2$/s is the diffusion constant,
$\mu = D/(k_B T)$ is the mobility of the particles,
$N[0,1]$ is a Gaussian distributed random number with mean of $0$ and
standard deviation of $1$, and $\Delta t = 1$ ms is the size of a simulation time step.

\begin{acknowledgments}

  We thank P. Tierno, A. Ortiz, and J. Loehr for useful discussions and
  for providing realistic parameters with regards to the experimentally feasible
  realization of colloidal spin ice. We gratefully acknowledge the support of the U.S.
  Department of Energy through the LANL/LDRD program for this work. This work was
  carried out under the auspices of the NNSA of the U.S. DoE at LANL under Contract No.DE-AC52-06NA25396.
\end{acknowledgments}

\subsection*{Author contributions}
A.L. performed the numerical calculations.  C.N. performed the analytical calculations.
A.L., C.N., C.J.O.R., and C.R. contributed to analysing the data and writing the paper.

\subsection*{Additional information}
           {\bf Competing financial interests:} The authors declare no competing financial
           interests.

\begin{figure*}[th!]
\includegraphics[width=.9 \textwidth]{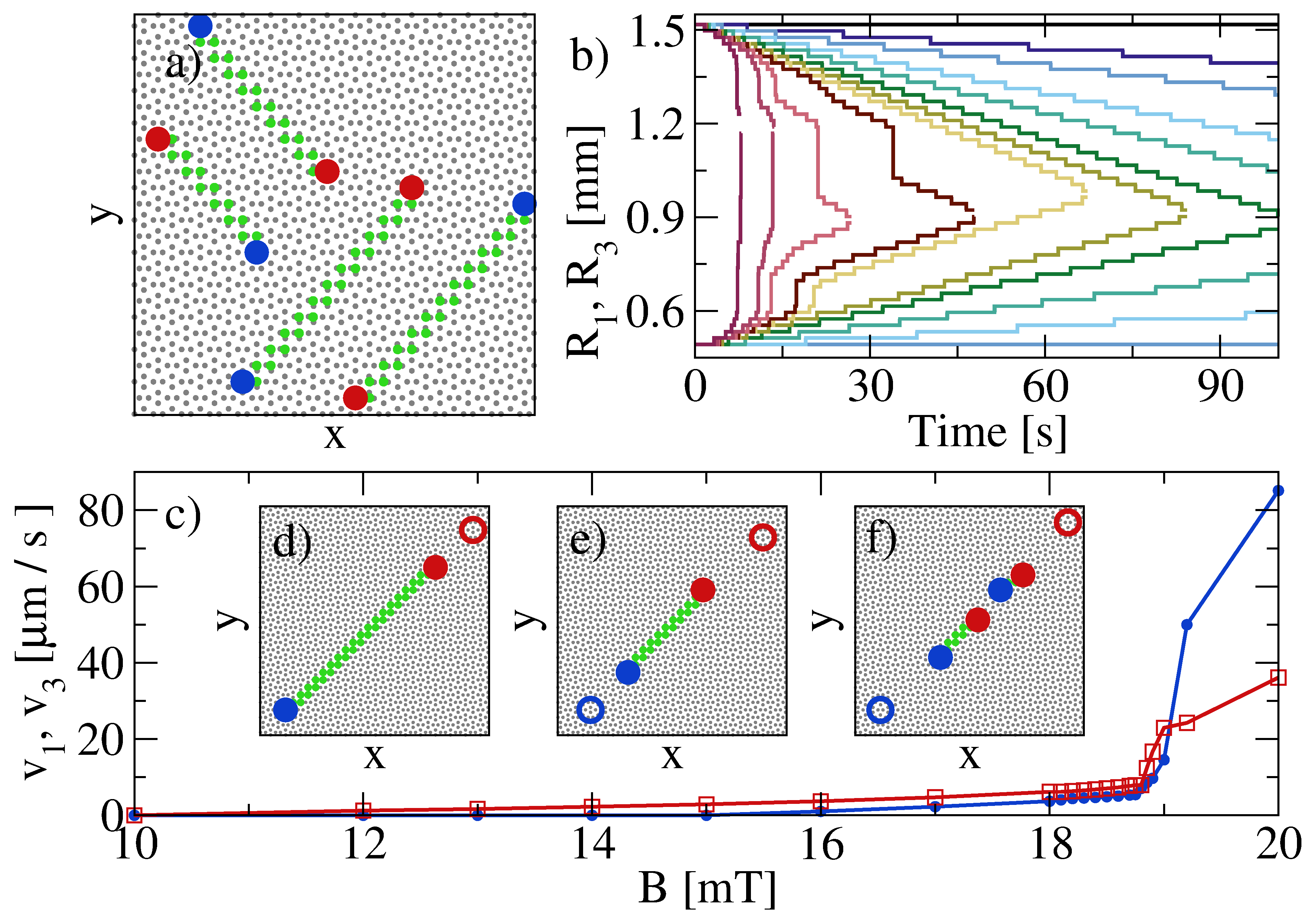}
\caption{
  Defect line images and motion.
  (a) The four possible arrangements of the defect lines in a portion of the sample.
  Red: 3-in vertex; blue: 1-in vertex; green: biased ground state vertex; gray: unbiased
  ground state vertex.  (b) The position $R_1$ of the 1-in (bottom lines) and
  $R_3$ of the 3-in (top lines)
  ends of a defect line vs time at magnetic fields
  $B=20$, 19.2, 19, 18.85, 18.8, 18.5, 18, 17, 16, 14, 12, and $10$ mT,
  from left to right.
  (c) The velocity $v_1$ (blue) and $v_3$ (red)
  of the defect ends calculated with a linear fit vs $B$.
  (d-f) Illustrations of the different modes of
  defect line contraction in a portion of the sample.
  Open circles indicate the original positions of
  the 3-in and 1-in ends, while closed circles show the final positions.
  (d) Contraction of only the 3-in end.
  (e) Contraction of both ends.  (f) Contraction of both ends accompanied by
  nucleation of new defect vertices along the defect line.  
}
\label{fig:2}
\end{figure*}

\begin{figure}
\includegraphics[width=\columnwidth]{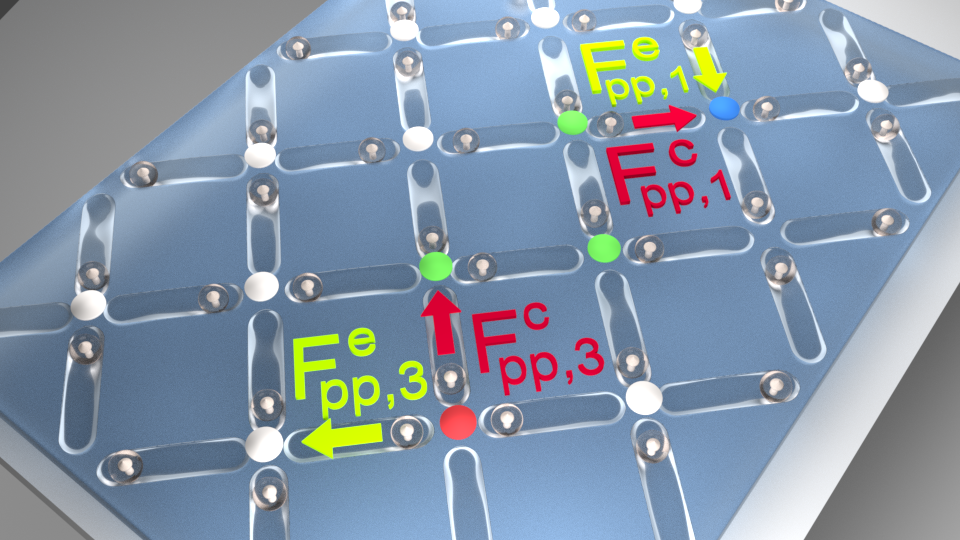}
\caption{
  Schematic showing the forces
  that are responsible for contracting and extending the defect line.
  The 1-in and 3-in ends of the line are marked blue and red, respectively,
  while the biased ground state vertices along the defect line are marked green.
  Particle-particle forces that act to extend (e, green lettering and arrows) or contract
  (c, red lettering and arrows) the defect are marked for the
  1-in end, $F_{pp,1}^e$ and $F_{pp,1}^c$, and for the 3-in end,
  $F_{pp,3}^e$,$F_{pp,3}^c$. 
}
\label{fig:3extra}
\end{figure}

\begin{figure*}
\includegraphics[width=  \textwidth]{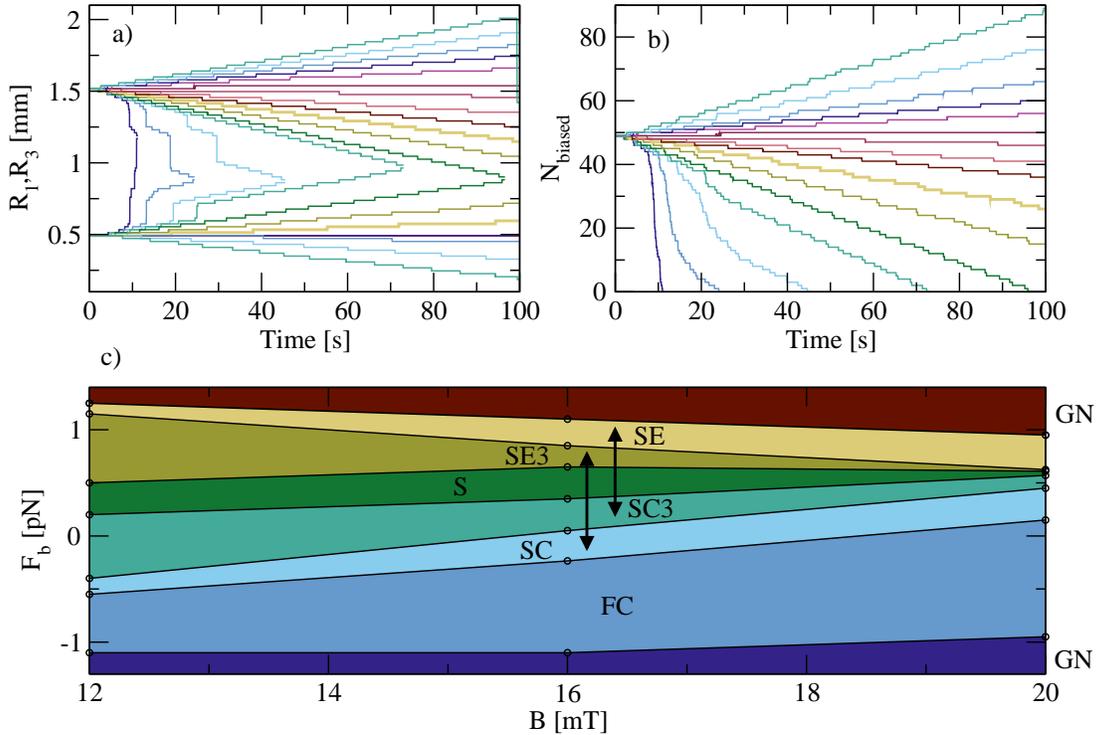}
\caption{
  Biased systems.
  (a) The positions $R_1$ (bottom lines) and $R_3$ (top lines) of the
  ends of a defect line
  vs time in a sample with $B = 16$mT for varied biasing fields 
  $F_b=-0.3$, -0.27, -0.25, -0.24, -0.2, -0.1, 0 (thick yellow line), 0.1, 0.2, 0.3, 0.5, 0.7, 0.8,
  0.9, 1.0, and 1.1, from left to right.
  (b) $N_{\rm biased}$, the number of vertices in the biased ground state,
  vs time in the same system for the same fields as in panel (a),
  $F_b=-0.3$, ... 1.1 from left to right.
  (c) Phase diagram as a function of $F_{\rm b}$ vs $B$
  showing the different phases of defect line contraction and expansion.
  Dark blue: Global nucleation of 1-in/3-in and biased ground state vertices (GN).
  Medium blue: Fast contraction with nucleation of 1-in/3-in vertex
  pairs along the defect line (FC).
  Light blue: Slow contraction on both ends of the defect line (SC).
  Light green: Slow contraction of only the 3-in end (SC3).
  Dark green: Stable defect string (S).
  Olive: Slow expansion of only the  3-in end (SE3).
  Yellow: Slow expansion on both ends of the line (SE).
  Red: Global nucleation of 1-in/3-in and biased ground state vertices (GN).
  The arrows indicate possible field combinations that can be applied in order to generate
  a forward or backward ratcheting defect line.
}
\label{fig:3}
\end{figure*}

\begin{figure*}
\includegraphics[width=\textwidth]{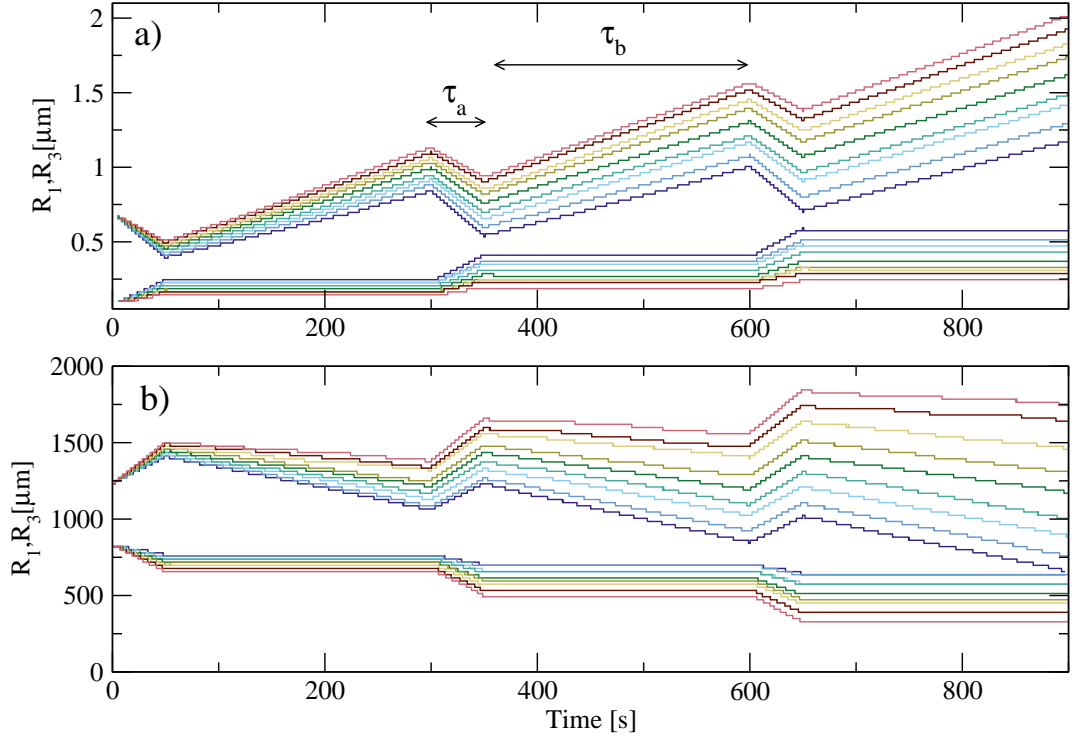}
\caption{Ratcheting defect lines.
  $R_1$ (bottom lines) and $R_3$ (top lines)
  vs time in
  samples with $B=16$mT for
  alternating drive intervals with bias
  $B_a$ applied for $\tau_a=50$ s and
  $B_b$ applied for $\tau_b=250$ s 
  during each cycle.
  (a) Forward ratchet effect
  for ($B_a$,$B_b$) values of $(-0.18,0.76)$, $(-0.16,0.77)$, $(-0.14,0.78)$, $(-0.12,0.79)$, 
  $(-0.10,0.8)$, $(-0.08,0.81)$, $(-0.06,0.82)$, $(-0.04,0.83)$
  and $(-0.02,0.84)$, from blue to red. (b) Reverse ratchet effect for
  ($B_a$,$B_b$) values of
  $(0.96,0.22)$, $(0.98,0.23)$, $(1.0, 0.24)$,
  $(1.02, 0.25)$ ,$(1.04, 0.26)$, $(1.06, 0.27)$, $(1.08, 0.28)$, $(1.1, 0.29)$     
  and $(1.12, 0.3)$, from blue to red.
}
\label{fig:4}
\end{figure*}

\end{document}